\documentclass[useAMS,usenatbib,usegraphicx]{mn2e}
\usepackage{amsmath} 
\usepackage{amsfonts}
\usepackage{amssymb}
\input epsf

\title[Magnetic field intensification in the Sun]{Convective intensification
of magnetic fields in the quiet Sun}
\author[P. J. Bushby et al.]{P. J. Bushby$^{1}$\thanks{E-mail: paul.bushby@ncl.ac.uk
(PJB)}, S. M. Houghton$^{2}$, M. R. E. Proctor$^{3}$ and
N. O. Weiss$^{3}$\\ $^{1}$School of Mathematics and Statistics, 
Newcastle University, Newcastle upon Tyne NE1 7RU, UK\\
$^{2}$Department of Applied Mathematics, University of Leeds, Leeds
LS2 9JT, UK\\
$^{3}$DAMTP, Centre for Mathematical Sciences,
University of Cambridge, Wilberforce Road, Cambridge CB3 0WA, UK}

\date{Submitted 08/01/08}

\pagerange{\pageref{firstpage}--\pageref{lastpage}}

\pubyear{2008}

\begin{document}

\maketitle

\label{firstpage}

\begin{abstract}
Kilogauss-strength magnetic fields are often observed in intergranular
lanes at the photosphere in the quiet Sun.  Such fields are stronger than
the equipartition field $B_e$, corresponding to a magnetic energy density
that matches the kinetic energy density of photospheric convection, and
comparable with the field $B_p$ that exerts a magnetic pressure equal to
the ambient gas pressure.  We present an idealised numerical model of
three-dimensional compressible  magnetoconvection at the photosphere,
for a range of values of the magnetic Reynolds number.  In the absence
of a magnetic field, the convection is highly supercritical and is
characterised by a pattern of vigorous, time-dependent, ``granular''
motions. When a weak magnetic field is imposed upon the convection, 
magnetic flux is swept into the convective downflows where it forms
localised concentrations. Unless this process is significantly inhibited
by magnetic diffusion, the resulting fields are often much greater than
$B_e$, and the high magnetic pressure in these flux elements leads to
their being partially evacuated. Some of these flux elements contain
ultra-intense magnetic fields that are significantly greater than $B_p$.
Such fields are contained by a combination of the thermal pressure of the
gas and the dynamic pressure of the convective motion, and they are
constantly evolving.  These ultra-intense fields develop owing to
nonlinear interactions between magnetic fields and convection; they
cannot be explained in terms of ``convective collapse'' within a thin flux
tube that remains in overall pressure equilibrium with its
surroundings.

\end{abstract}

\begin{keywords}
convection -- (magnetohydrodynamics:) MHD -- 
Sun: granulation -- Sun: magnetic fields --
Sun: photosphere
\end{keywords}

\section{Introduction}

High resolution observations of the solar photosphere have greatly
enhanced our understanding of the complex interactions between
magnetic fields and turbulent convective motions. Modern ground-based
telescopes, such as the SST on La Palma, and space-borne instruments,
such as the Solar Optical Telescope (SOT) on the recently-launched
Hinode satellite, can now resolve the fine structure of photospheric
magnetic fields. Such high resolution observations have already
provided new insights into the magnetic field structures that can be
found in sunspot penumbrae (Scharmer et al. 2002), plage regions
(Berger et al. 2004) and the quiet Sun (see, for example, Centeno et
al. 2007; Rezaei et al. 2007). In the quiet Sun, localised
concentrations of intense vertical magnetic flux often accumulate in
the convective downflows (Lin \& Rimmele 1999; Dom\'inguez Cerde\~na,
Kneer \& S\'anchez Almeida 2003; Centeno et al. 2007). These localised
flux concentrations typically show up as bright points in G-band
images of the photospheric granulation (see, for example, Berger \&
Title 1996, 2001). Direct measurements indicate that the peak field
strength within these small-scale magnetic features is often well in
excess of a kilogauss (see, for example, Grossmann-Doerth, Keller \&
Sch\"ussler 1996; Dom\'inguez Cerde\~na et al. 2003). Although
kilogauss-strength magnetic features occupy only a small fraction of
the quiet solar surface, these localised flux concentrations contain a
significant proportion of the total (unsigned) quiet Sun magnetic flux
(Dom\'inguez Cerde\~na, S\'anchez Almeida \& Kneer 2006; S\'anchez
Almeida 2007). 

\par The origin of quiet Sun magnetic fields remains an open
question. Either these magnetic fields are generated locally at the
photosphere, as the result of small-scale convectively-driven dynamo
action (Cattaneo 1999; Cattaneo \& Hughes 2006; V\"ogler \& Sch\"ussler
2007), or they were originally generated elsewhere in the solar
interior and are simply being re-processed and amplified by the local convective
motions. What is clear, however, is that the granular convective
upwellings at the solar photosphere will tend to expel magnetic flux,
causing it to accumulate in the convective downflows (Proctor \& Weiss
1982). This explains the observed association between quiet Sun
magnetic fields and the intergranular lanes. However, the peak
field strength that is measured in these regions is more difficult to
explain. Estimates of the kinetic energy density of the granular
convection in the solar photosphere suggest that magnetic field
strengths greater than approximately $400$G would exceed the
``equipartition'' value, $B_e$, at which the magnetic energy density
of these localised fields is comparable to the granular convective
kinetic energy density (see, for example, Galloway, Proctor \& Weiss
1977). Since the magnetic energy density is proportional to the square
of the magnetic field strength, the observed kilogauss-strength fields
in the quiet Sun have a magnetic energy density that is an order of
magnitude larger than equipartition. In fact, a better estimate for
the observed field strengths is given by $B_p$, which corresponds to
the field strength at which the local magnetic pressure balances the
ambient granular gas pressure.  

\par In theoretical studies of Boussinesq magnetoconvection (see e.g. Galloway
et al. 1977; Galloway, Proctor \& Weiss 1978; Galloway \& Moore 1979),
the process of magnetic flux expulsion leads naturally to the formation of
localised flux concentrations at the edges of convective cells. For
axisymmetric converging flows, the {\it kinematic} flux concentration of
weak magnetic fields produces peak fields that scale linearly with the
magnetic Reynolds number, $Rm$, whilst the width of the resulting
magnetic feature is proportional to $Rm^{-1/2}$. However, for stronger
fields this amplification process is limited by {\it dynamical}
effects. Since the magnetic pressure does not play a role in Boussinesq
magnetoconvection, any magnetic feedback is due to the non-conservative
component of the Lorentz force, which results from the magnetic
curvature. In the dynamical regime, the peak field can only exceed
$B_e$ for large values of the magnetic Prandtl number (the ratio of
the kinematic viscosity of the fluid to its magnetic
diffusivity). Formally, this scaling can only be verified for small
Reynolds numbers, $Re$, although it appears not to be restricted to this
parameter regime (Cattaneo 1999). Kerswell \& Childress (1992) also
found a similar magnetic Prandtl number dependence in their idealised
boundary-layer study of the equilibrium of a thin flux tube in steady
compressible convection (see also Cameron \& Galloway
2005). Unfortunately, estimates of the magnetic Prandtl number in the
solar photosphere indicate that this diffusivity ratio is extremely
small (see, for example, Ossendrijver 2003). So although these
comparatively idealised models can explain the formation of highly
localised magnetic features at large $Rm$, they are unable to account
for the appearance of super-equipartition, kilogauss-strength magnetic
fields in the quiet Sun.  
      
\par Since the peak magnetic field in kilogauss-strength
magnetic features is comparable to $B_p$, the magnetic pressure is
bound to play a significant role in the local dynamics. If a magnetic
feature is assumed to be in pressure balance with its non-magnetic
surroundings, the internal gas pressure must be smaller than that of
the surrounding fluid in order to compensate for the large
magnetic pressure. Unless these features are much cooler than their
surroundings, this reduction in gas pressure is most easily achieved
if these magnetic regions are at least partially evacuated (see, for
example, Proctor 1983; Proctor \& Weiss 1984). This effect is absent
from Boussinesq models, which neglect the effects of compressibility,
so it is inevitable that such models underestimate the peak fields
that can be generated by the process of flux concentration at the
solar photosphere. When seeking to explain the formation of
kilogauss-strength magnetic fields in the quiet Sun, it is therefore
important to consider the effects of compressibility (see e.g. Hughes
\& Proctor 1988). 

\par The most widely studied compressible model of magnetic field
amplification in the quiet Sun is the ``convective collapse''
instability of thin magnetic flux tubes (Webb \& Roberts
1978; Spruit 1979; Spruit \& Zweibel 1979; Unno \& Ando
1979). Convective collapse models consider the linear stability of a
thin vertical magnetic flux tube embedded in a non-magnetic,
superadiabatically stratified atmosphere. There are initially no
convective motions along the tube, which is assumed to be in
pressure balance with its surroundings. Provided that the initial
magnetic field is not too strong, the flux tube is subject to a
convective instability that drains plasma vertically out of the
tube. As the plasma drains out of the tube, the local pressure balance
implies that the tube must ``collapse'' to form a narrower (and therefore more
intense) magnetic flux concentration. This instability will continue
to operate until the magnetic field is strong enough to suppress
convective motions. This model certainly represents a plausible
mechanism for the formation of kilogauss-strength magnetic
fields (the only restriction being that the peak field, $B_{max} \le
B_p$). Having said that, this model has its limitations
(Hughes \& Proctor 1988; Thomas \& Weiss 1992, 2008). In
particular, the thin flux tube approximation does not appear to be
consistent with photospheric observations (see, e.g. Berger \& Title
1996; Berger et al. 2004): even in the quiet Sun, the observed
magnetic regions seem to form a constantly-evolving ``magnetic fluid''
(see, for example, Thomas \& Weiss 2008), and so cannot simply be regarded
as a collection of discrete collapsed flux tubes. In addition, the
static equilibrium for the instability is rarely achievable in photospheric
magnetoconvection. Finally, by reducing the horizontal component of
the momentum equation to a simple pressure balance, the model not only
ignores the possible dynamical influence of the surrounding convective
motions, but also neglects the possible effects of magnetic curvature,
which may play an important role in ensuring the equilibrium of
more extensive magnetic features (see, for example, V\"ogler et
al. 2005, where curvature effects help to support a peak field,
$B_{max}$, in excess of $B_p$).   

\par More realistic models of this process can only be studied by
carrying out large-scale numerical simulations of convective magnetic
flux intensification. In order to make detailed comparisons between numerical
simulations and spectral observations, it is necessary to take account
of effects such as partial ionisation and radiative transfer (see,
e.g., Nordlund 1982; V\"ogler et al. 2005). This approach has
successfully simulated magnetic fields and spectral features that are
similar to those observed in plages (Grossmann-Doerth, Sch\"ussler \&
Steiner 1998; Keller et al. 2004; Carlsson et al. 2004; V\"ogler et
al. 2005), in sunspot umbrae (Sch\"ussler \& V\"ogler 2006), and in
the quiet Sun (Khomenko et al. 2005; Stein \& Nordlund 2006). More
idealised models of photospheric magnetoconvection focus entirely upon
the interactions between compressible convection and magnetic fields
(e.g. Hurlburt \& Toomre 1988; Matthews, Proctor \& Weiss 1995; Weiss
et al. 1996; Rucklidge et al. 2000; Weiss, Proctor \& Brownjohn 2002;
Bushby \& Houghton 2005). This complementary approach lends itself
more easily to systematic surveys of parameter space, and has had
considerable success in qualitatively reproducing solar-like
behaviour. 

\par In this paper, we investigate the formation of
super-equipartition, localised magnetic features in the quiet Sun. We
carry out numerical simulations of an idealised model of photospheric
magnetoconvection. Clearly it is not possible to achieve solar-like
values of the viscous and magnetic Reynolds numbers in these
simulations (which are both very large at the solar photosphere, see
e.g. Ossendrijver 2003). However the Reynolds numbers that are used
here are large enough that these simulations (at least qualitatively)
reproduce most of the key physical features of flux intensification in
photospheric magnetoconvection. The set-up of this model and the
numerical results that are obtained from it are described in the next
two Sections of the paper. In the final part, we discuss the relevance
of these results to photospheric magnetic fields and compare our
findings with predictions of the simplified ``convective collapse''
models.

\section{Governing equations and numerical methods}

The model that is considered in this paper is an idealised, local
representation of magnetoconvection in the quiet Sun. We
solve the equations of three-dimensional compressible
magnetohydrodynamics for a plane layer of electrically-conducting,
perfect monatomic gas. The layer of gas is heated from below and
cooled from above. We assume constant values for the gravitational
acceleration $g$, the shear viscosity $\mu$, the magnetic diffusivity
$\eta$, the thermal conductivity $K$, the magnetic permeability
$\mu_0$ and the specific heat capacities at constant density and
pressure ($c_v$ and $c_p$ respectively). The axes of the chosen
Cartesian frame of reference are oriented so that the $z$-axis
points vertically downwards (parallel to the gravitational
acceleration). Defining $d$ to be the depth of the convective layer,
the gas occupies the region $0 \le x, y \le 4d$ and $0 \le z \le d$,
which gives a wide Cartesian domain with a square horizontal
cross-sectional area. Periodic boundary conditions are imposed in each
of the horizontal directions. Idealised boundary conditions are
imposed at $z=0$ (the upper surface) and $z=d$ (the lower
surface). These bounding surfaces are held at fixed temperature, and
are assumed to be impermeable and stress-free. In addition, any
magnetic field that is present is constrained to be vertical at the
upper and lower boundaries. Similar models have been considered in
several previous studies (see, for example, Matthews et al. 1995;
Rucklidge et al. 2000; Weiss et al. 2002; Bushby \& Houghton 2005). 

\par It is convenient to formulate the governing equations for
magnetoconvection in terms of non-dimensional variables. We adopt
non-dimensionalising scalings that are similar to those described by
Matthews et al. (1995). All lengths are scaled in terms of the depth
of the Cartesian domain, $d$, whilst the temperature, $T$, and
density, $\rho$, are both scaled in terms of their values at the upper
surface of the domain (which are denoted by $T_0$ and $\rho_0$
respectively). Defining $R_*$ to be the gas constant, the velocity,
$\mathbf{u}$, is scaled in terms of the isothermal sound speed at the top of
the layer, $\left(R_*T_0\right)^{1/2}$; a natural scaling for time is
therefore $d/\left(R_*T_0\right)^{1/2}$, which corresponds to an
acoustic timescale.  The parameter $\theta$ denotes the
(dimensionless) temperature difference between the upper and lower
boundaries. In the absence of any motion the gas is a polytrope with
a polytropic index  $m=gd/R_*T_0\theta -1$. 
Rather than relating the magnetic field strength
to the Chandrasekhar number (see, for example, Weiss et al. 2002;
Bushby \& Houghton 2005), we also scale the Alfv\'en speed at the top
of the layer in terms of the sound speed. This implies that the
appropriate non-dimensionalising scaling for the magnetic field,
$\mathbf{B}$, is given by
$\left(\mu_0\rho_0R_*T_0\right)^{1/2}$. Having made these scalings,
the governing equations for the density, the momentum density ($\rho
\mathbf{u}$), the magnetic field and the temperature are given by

\begin{eqnarray}
&&\frac{\partial \rho}{\partial t}=- \nabla \cdot \left(\rho
\bmath{u}\right),\\ \nonumber \\ 
&&\frac{\partial}{\partial t}\left(\rho \bmath{u}\right)=- \nabla
\left(P + |\bmath{B}|^2/2\right) +\theta(m+1)\rho\bmath{\hat{z}}\\
\nonumber&& \hspace{0.65in} + \nabla \cdot \left( \bmath{BB} - \rho \bmath{uu} +
\kappa \sigma \bmath{\tau}\right), \hspace{0.4in} \\ \nonumber \\
&&\frac{\partial \bmath{B}}{\partial t}=\nabla \times \left( \bmath{u}
\times \bmath{B} -  \kappa \zeta_0 \nabla \times \bmath{B} \right),
~~\nabla \cdot \bmath{B} = 0,\\ \nonumber \\
&&\frac{\partial T}{\partial t}= -\bmath{u}\cdot\nabla T -
\left(\gamma -1\right)T\nabla \cdot \bmath{u} +
\frac{\kappa\gamma}{\rho}\nabla^2 T \\ \nonumber &&\hspace{0.35in}+
\frac{\kappa(\gamma-1)}{\rho}\left(\sigma \tau^2/2 + \zeta_0|\nabla
\times \bmath{B}|^2\right).
\end{eqnarray}

\noindent The components of the stress tensor, $\bmath{\tau}$ are
given by

\begin{equation}
\tau_{ij}= \frac{\partial u_i}{\partial x_j}+\frac{\partial
 u_j}{\partial x_i} - \frac{2}{3}\frac{\partial u_k}{\partial
 x_k}\delta_{ij},
\end{equation}

\noindent whilst the pressure, $P$ is determined by the equation of
state for a perfect gas

\begin{equation}
P=\rho T.
\end{equation}

\par These governing equations are characterised by several
non-dimensional constants, including the Prandtl
number, $\sigma=\mu c_p/K$, the (non-dimensional) thermal diffusivity,
$\kappa=K/\rho_0 d c_p \left(R_*T_0\right)^{1/2}$, and the ratio of
specific heats, $\gamma=c_p/c_v$. The ratio of the magnetic
diffusivity to the thermal diffusivity, $\zeta$, is proportional to
the fluid density $\rho$ (and therefore increases with depth). At
the top of the layer $\zeta=\zeta_0\equiv\eta
\rho_0 c_p/K$. If all the other parameters are fixed, varying $\kappa$
is equivalent to varying the mid-layer Rayleigh number, 

\begin{equation}
Ra=\left(m+1-m\gamma\right)\left(1+\theta/2\right)^{2m-1}\frac{(m+1)\theta^2}{\kappa^2\gamma\sigma}.
\end{equation}

\noindent This Rayleigh number measures the destabilising effects of a
superadiabatic temperature gradient relative to the stabilising
effects of diffusion. The parameters are all described in greater
detail by Matthews et al. (1995). 

\par The simplest non-trivial equilibrium solution of these governing
equations corresponds to a static polytrope with a uniform magnetic
field. Initially we restrict attention to the case in which
$\mathbf{B}=0$. In this equilibrium solution, $\mathbf{u}=0$,
$T(z)=1+\theta z$ and $\rho(z)=\left(1+\theta z\right)^m$. Fixing
$\theta=10$ and $m=1$ gives a highly stratified atmosphere in which
the temperature and density both vary by an order of magnitude across
the layer. We consider a monatomic gas, therefore $\gamma=5/3$. This
implies that the $m=1$ polytrope is superadiabatically stratified. For
simplicity (and for ease of comparison with previous studies) we set
the Prandtl number equal to unity, i.e. $\sigma=1$. A range of values
for $\zeta_0$ is considered in this paper. Finally, the Rayleigh
number is chosen to be $Ra=4.0 \times 10^5$. This value for $Ra$ is
more than two orders of magnitude larger than the critical value for
the onset of convective instabilities. 

\par The model described above is investigated by carrying out large-scale
numerical simulations. For this $4 \times 4 \times 1$ Cartesian
domain, we adopt a computational mesh of $256 \times 256 \times 160$
grid points. The horizontal extent of this computational
domain is smaller than in some of our previous calculations (see,
e.g. Bushby \& Houghton 2005). This reduction in box size does
influence the global structure of the convective motions by
eliminating mesoscale structures (Rucklidge et al. 2000). However, the
horizontal extent of the computational domain probably has little
influence upon the localised process of flux concentration. The
advantage of considering a smaller domain is that it allows us to
carry out high resolution numerical simulations without requiring an
excessive number of grid points. This enables us to model
magnetoconvective behaviour more accurately at high Reynolds numbers,
so this reduction in box size is a reasonable compromise. We use a
well tested code (see, e.g. Matthews et al. 1995) in which horizontal
derivatives are evaluated in Fourier space, whilst vertical
derivatives are evaluated using fourth order finite differences. The
time-stepping is carried out via an explicit 3rd order Adams-Bashforth
scheme, with a variable time-step. The code is efficiently
parallelised using MPI, and these simulations have made use of the
Cambridge-Cranfield High Performance Computing Facility and the UKMHD
Consortium machine based in St Andrews.  

\begin{figure}
\begin{center}
\epsfxsize\hsize\epsffile{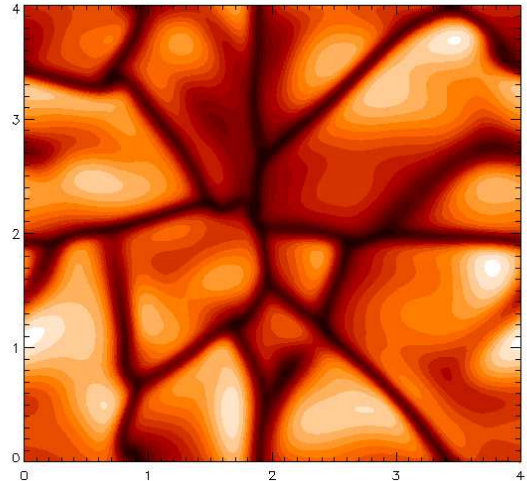}
\caption{The temperature distribution for non-magnetic convection, in
  a horizontal plane near the upper surface of the Cartesian
  domain, at $(z/d)=0.05$. Temperature contours are evenly
  spaced between $T=1.4$ and $T=2.6$ (the same contour spacings are
  used for all temperature plots in this paper). Brighter regions correspond to
  warmer fluid, cooler areas of fluid are represented by darker
  regions.\label{fig:1}}  
\end{center}
\end{figure}

\section{Results}

\subsection{The initial state}

The starting point for these idealised simulations is fully developed
non-magnetic convection. In order to generate such a convective state,
we consider the equilibrium solution corresponding to a static
polytrope with no applied magnetic field, and introduce a small
amplitude random temperature perturbation. This convectively-unstable
configuration is allowed to evolve until it has relaxed to a
statistically steady hydrodynamic state, as illustrated in
Fig.~\ref{fig:1}, which shows a snapshot of the temperature
distribution in a horizontal plane near the upper surface of the
computational domain. In this granular pattern, bright regions
correspond to broad warm upflows, whilst darker regions correspond to
cool narrow downflows. The strength of the convection can be
characterised by the mid-layer Reynolds number $Re=\rho_{\rm
  mid}U_{\rm rms}d/\mu$, where $U_{\rm rms}$ corresponds to the
rms-velocity of the convection and $\rho_{\rm mid}$ is the density at
the mid-layer of the original static polytrope. The Reynolds number
here is approximately $Re=150$. As noted in the Introduction,
realistic Reynolds numbers for photospheric convection are numerically
unobtainable, but the Reynolds number is large enough that
instructive results can be obtained in these idealised
calculations. 

\par Having established this (purely hydrodynamic) convective state,
we then introduce a weak uniform vertical magnetic field,
$B_0\bmath{\hat z}$, with $B_0$ chosen so that the initial magnetic
energy is approximately 0.1\% of the kinetic energy. A weak imposed
field of this form tends to favour the formation of highly localised
magnetic features (see, for example, Bushby \& Houghton 2005). In what
follows, the time at which this magnetic field is introduced is
denoted by $t=0$. The subsequent evolution of this magnetic field
depends crucially upon the magnetic Reynolds number of the flow,
$Rm=U_{\rm rms}d/\eta$. In non-dimensional terms, $Rm \propto
\zeta_0^{-1}$; thus it is possible to investigate a range of
values of $Rm$ simply by repeating the numerical experiment with
different values of $\zeta_0$. The range investigated is
$0.2\leq\zeta_0\leq 2.4$ ($120 \gtrsim Rm \gtrsim 10)$. Larger values of
$Rm$ correspond to less diffusive plasmas, and are therefore more
relevant to photospheric magnetoconvection, although (as with $Re$) it
is not yet possible to carry out fully resolved simulations with
realistic values of $Rm$ for photospheric magnetoconvection. We
therefore focus initially upon the case of $\zeta_0=0.2$
($Rm\simeq 120$). The effects of varying $\zeta_0$ are discussed later in
this Section.

\begin{figure}
\begin{center}
\epsfxsize\hsize\epsffile{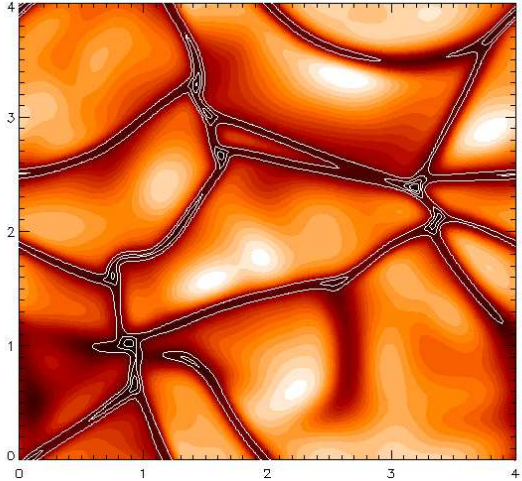}
\caption{The magnetic field distribution during the flux expulsion
  phase, at $t=0.24$. The shaded contours show the temperature distribution
  in a horizontal plane at $(z/d)=0.05$ (as described in
  Fig.~\ref{fig:1}). The white contours denote constant values of the
  vertical component of the magnetic field.\label{fig:2}}
\end{center}
\end{figure}

\subsection{Results for $\zeta_0=0.2$}

Since the initial magnetic field is comparatively weak, the Lorentz
forces play a minor role in the very early stages of evolution in
these numerical simulations. During this brief ``kinematic'' phase,
diverging convective motions at the upper surface of the computational
domain rapidly expel magnetic flux from the granular interiors. This
process causes magnetic flux to accumulate in the convective downflows
in the intergranular lanes. The strongest field concentrations tend to
occur at the vertices between neighbouring granules, due to
the fact that any flows along the intergranular lanes tend to converge
upon these vertices. The process of flux expulsion and accumulation
is illustrated in Fig.~\ref{fig:2}, which shows the distribution of
the vertical component of the magnetic field during this flux
expulsion phase. In this (relatively) high $Rm$ regime, the width of
the localised magnetic features is comparable to the width of the
narrow intergranular lanes. 

\par As these localised concentrations of magnetic flux form, the
principle of flux conservation implies that the magnetic energy
density in these regions rapidly increases to the point where the
Lorentz forces become dynamically significant. In the upper layers of
the computational domain, most of the magnetic energy resides in the
vertical component of the magnetic field, and the horizontal gradients
in this field component are usually much larger than any variations in
the vertical direction. This implies that the horizontal magnetic
pressure gradient (i.e. the horizontal component of
$\nabla\left[\mathbf{B}^2/2\mu_0\right]$) is the dominant component of
the Lorentz force. The effects of the magnetic pressure are most
apparent near the surface, where the gas pressure is relatively
small. The magnetic pressure gradient tends to inhibit the converging
convective motions that are responsible for driving the flux
concentration process. However, the flux amplification process is not
immediately suppressed, because the total (i.e. gas plus magnetic)
pressure increases more gradually than the magnetic pressure
alone. This is due to the fact that the local convective downflows
rapidly carry fluid away from the surface regions of the magnetic
feature. These downflows are responsible for partially evacuating the
upper regions of the magnetic features, which in turn leads to a
reduction in the local gas pressure. This reduction in gas pressure
(at least partially) compensates for the increased magnetic
pressure. In the most intense magnetic flux concentrations, the gas
pressure drops to as little as a few percent of its initial value at
the surface before the magnetic field becomes locally strong enough to
suppress the vertical convective motions. A combination of the
suppression of these vertical motions plus the effects of diffusion
eventually halts this flux concentration process. 

\begin{figure*}
\begin{center}
\epsfxsize\hsize\epsffile{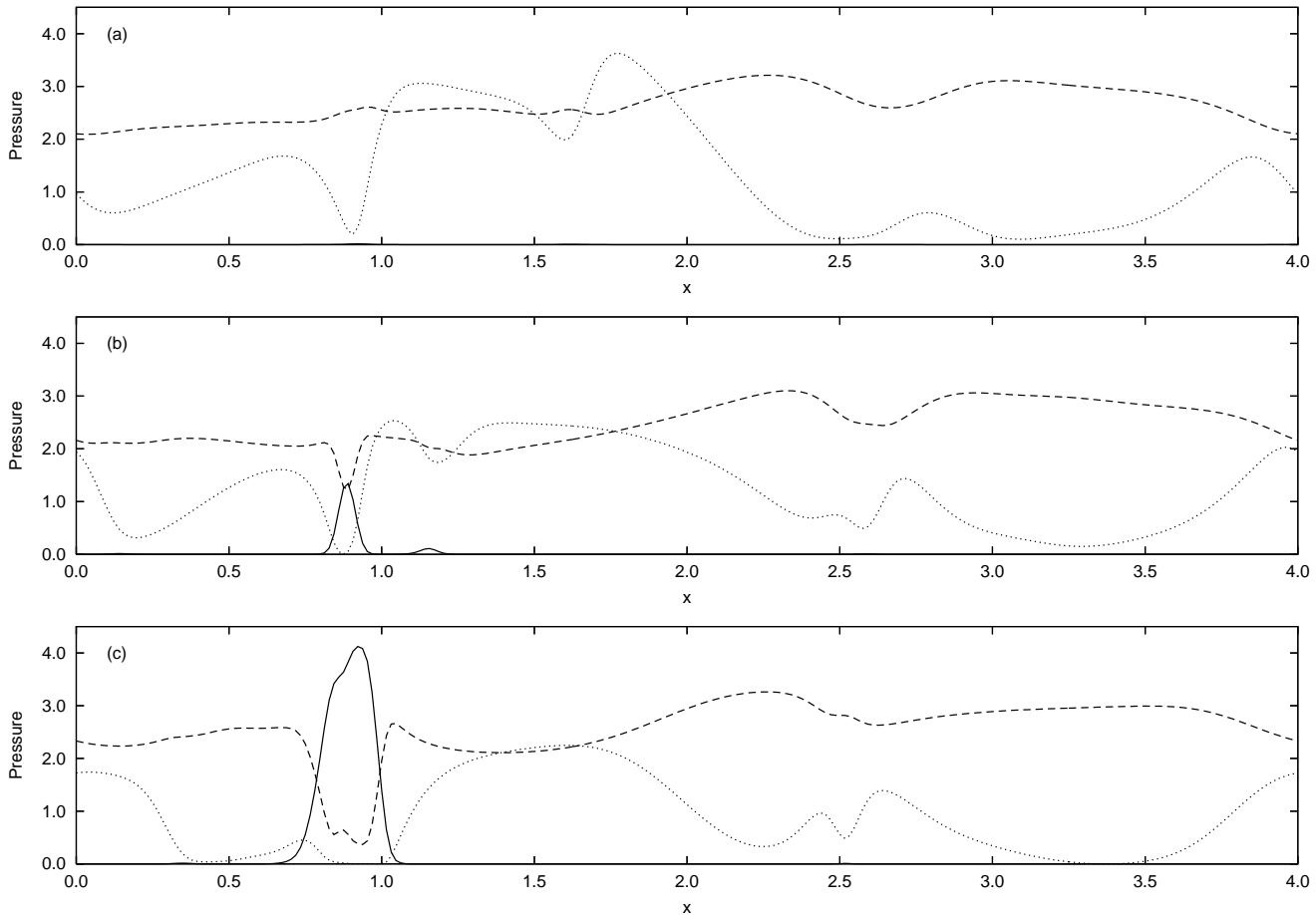}
\caption{The $x$-dependence of the pressure distributions (each at
  fixed $y$) at three different time intervals along a horizontal cut
  through the strongest magnetic feature at the upper surface. Solid
  lines correspond to the magnetic pressure, $P_{\rm mag}$; dashed lines
  correspond to the gas pressure, $P_{\rm gas}$; the dotted lines
  represent the dynamic pressure, $P_{\rm dyn}$ (see text). These
  snapshots are taken at (a) t=0.12, (b) t=0.61 and (c)
  t=1.61.\label{fig:3}}
\end{center}
\end{figure*}

\par To describe this process in a more quantitative fashion, it
is useful not only to consider the variations of the gas pressure
($P_{\rm gas}$) and the magnetic pressure ($P_{\rm mag}$), but also the
dynamic pressure ($P_{\rm dyn}$), which represents the
dynamical influence of the convective motions. Following Hurlburt \&
Toomre (1988), we define $P_{\rm dyn}=\rho|\mathbf{u}|^2$ (although note
that other definitions have been used in some other studies, e.g.,
Weiss et al. 1996). Whilst this expression does not correspond
directly to a pressure term in the momentum equation $(2)$, it does
usefully quantify the vigour of the convective motions. Since the peak
flow speeds at the surface are comparable to the local sound speed
(any shocks being smoothed out by the viscosity of the fluid), we
would expect $P_{\rm dyn}$ to play a significant dynamical
role. Fig.~3 illustrates the time-evolution of these pressure
distributions at the surface (where the most significant variations
are observed) immediately after the magnetic field is introduced. To
generate a one-dimensional pressure map, we fix the value of $y$ so
that a horizontal cut (in the $x$ direction) along the upper surface
of the computational domain passes through the strongest magnetic
feature. The three plots in Fig.~3 show the surface pressure
distributions along such a cut at three different times. These plots
clearly illustrate the scenario that was described in the previous
paragraph. As the magnetic pressure grows, there is a corresponding
decrease in the gas pressure as fluid rapidly drains out of the
surface regions of the magnetic feature. Once the feature has formed
(lower plot of Fig.~3), convective motions are strongly suppressed and
the gas pressure in the magnetic region is very much smaller than that
of the surrounding field-free fluid. Although partial evacuation,
with accompanying field intensification, has
already been observed in several previous studies (e.g. Hurlburt \&
Toomre 1988; Weiss et al. 1996; V\"ogler et al. 2005), the level of
evacuation is much more dramatic in these simulations.

\par The most surprising aspect of Fig.~3 is the fact that the
magnetic pressure within the upper layers of the resulting magnetic flux
concentration is much larger than the gas pressure of the surrounding
non-magnetic fluid. Since the ambient gas pressure increases rapidly with
depth ($P_{\rm gas} \propto (1+10z)^2$ in the unperturbed polytropic
atmosphere), the difference between the surrounding gas pressure and the
internal magnetic pressure decreases as we move away from the
surface. For this magnetic feature, the magnetic pressure is
comparable to the external gas pressure at a depth of approximately
$z=0.05$. However, even down to depths of approximately $z=0.1$, the sum
of the internal gas and magnetic pressures is still larger than the
external gas pressure. So, even below the surface, this magnetic
feature is not simply in pressure balance with its non-magnetic
surroundings. V\"ogler et al. (2005) have shown that magnetic
curvature effects can play a key role in the confinement of magnetic
flux concentrations. However, in this case the field lines are
predominantly vertical near the upper surface, so magnetic curvature
effects are negligible. Note that the lack of field-line curvature
also implies that these magnetic features are not being confined in
deeper layers (where the gas pressure is higher), since this would
require the flux concentrations to spread out in the upper layers. 

\par In the absence of a confining influence due to magnetic
curvature, the observed pressure imbalance implies
that the surrounding convective motions must be playing an active role
in the confinement of this magnetic feature. Put differently, where
the flow converges on an intense flux concentration there has to be an
excess stagnation pressure. Here this excess pressure is provided by
the magnetic field in a region where the gas pressure is reduced. This
is in total contrast with the simulation of umbral convection by
Sch\"ussler and V\"ogler (2006), where a divergent rising plume is
associated with a weaker field, and the excess pressure results
instead from an enhancement of density (associated with buoyancy
braking). The significance of the dynamic pressure in these
calculations is illustrated in Fig.~3 -- it is clear that the dynamic
pressure around the magnetic feature can often be comparable to the
local gas pressure. This shows that the dynamical influence of the
surrounding convective motions cannot be ignored when considering
models of photospheric magnetic field intensification. It should also be
stressed that the magnetic energy density of the flux concentration
that is illustrated in Fig.~3 is much larger than the mean kinetic
energy density (i.e. $P_{\rm dyn}/2$) of the surrounding granular
convection. These simulations provide confirmation of the fact that
the process of partial evacuation is important in the production of
such super-equipartition fields. Without such a reduction in the local
gas pressure, it is difficult to see how convective motions alone
could produce such a strong magnetic feature. 

\begin{figure}
\epsfxsize\hsize\epsffile{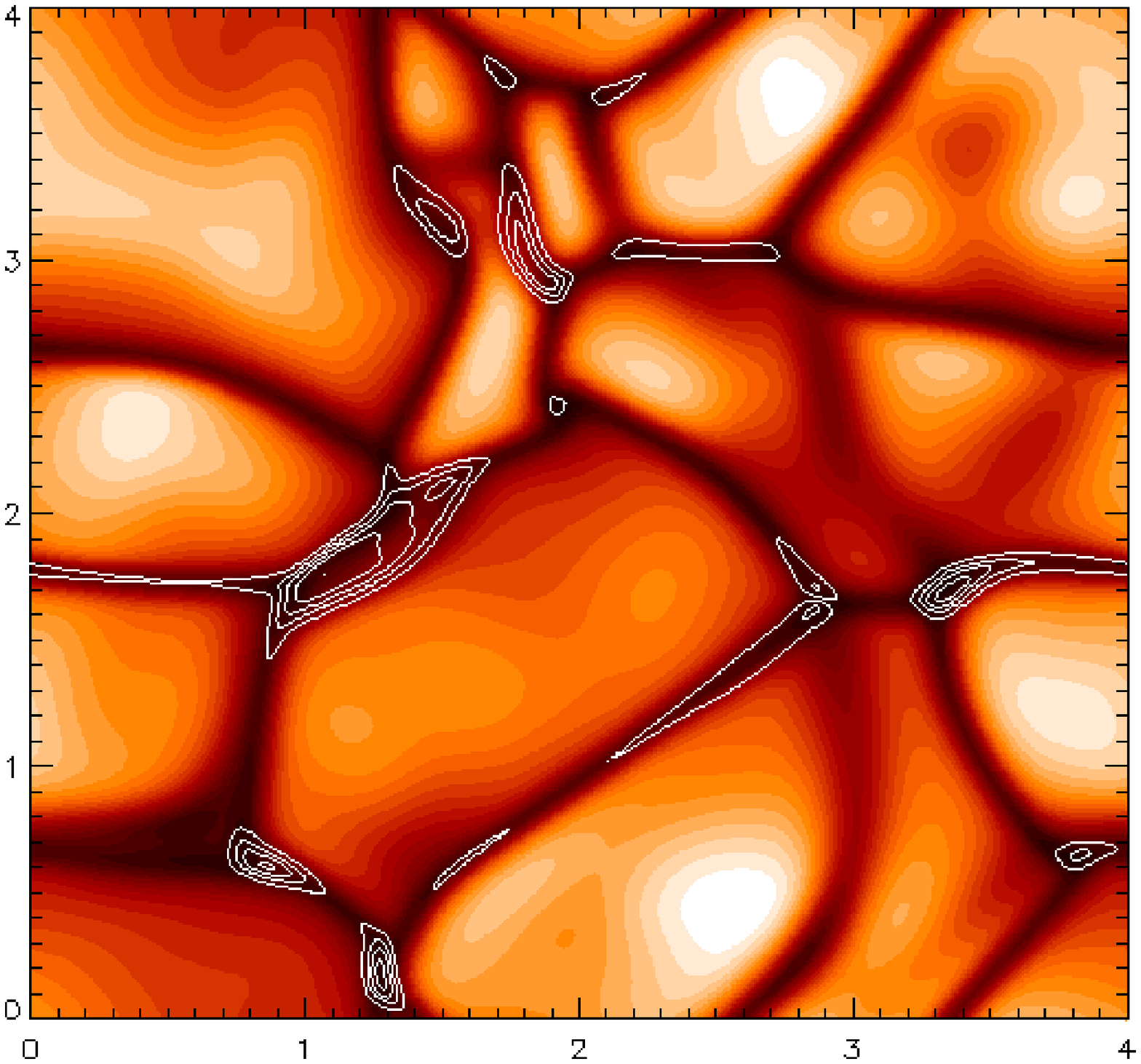}
\vspace{0.2cm}
\epsfxsize\hsize\epsffile{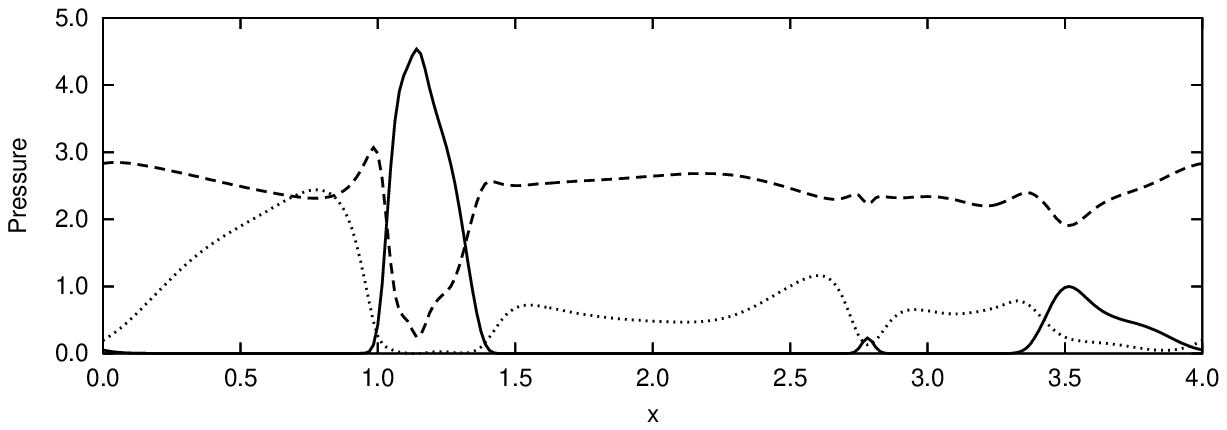}
\caption{The magnetic field distribution at $t=26.26$ (after several
  convective turnover times). Top:
  Like Fig.~2, this shows contours of the vertical magnetic field
  component superimposed upon the temperature profile in a horizontal
  plane at $(z/d)=0.05$. Bottom: The pressure
  distributions along a horizontal cut through the strongest magnetic
  feature at the surface. As in Fig.~3, the solid line corresponds to
  the magnetic pressure, the dashed line corresponds to the gas
  pressure, whilst the dynamic pressure is represented by a dotted
  line.\label{fig4}} 
\end{figure}
 
\par All the discussion so far has focused upon the early stages of
these numerical simulations. The process of magnetic field
intensification is arguably the most interesting phase of the
dynamics, although the evolution of the resulting magnetic features
also raises important issues. In the present idealised model, the net imposed
vertical magnetic flux is a conserved quantity, so there will always
be some accumulations of vertical magnetic flux somewhere within the
computational domain. However, what is less clear is whether or not
the occurrence of super-equipartition magnetic fields is a transient
feature of the model. Fig.~4 shows the magnetic field distribution
several convective turnover times after the magnetic field was first
introduced. It is clear that the most intense flux concentrations are
still partially evacuated, and it is also clear that the magnetic
energy density of these regions still exceeds the mean kinetic energy
density of the surrounding convection. In addition, the fact that the
internal magnetic pressure (in the upper layers) is still much larger
than the external gas pressure indicates that the surrounding
convective motions are still playing a key role in the confinement of
the magnetic feature. During their evolution, these features
continuously interact with the convective motions, which implies that
they are deformed, shredded and advected around the domain in a
time-dependent fashion. However, despite these complex interactions,
ultra-intense super-equipartition magnetic features seem to be a
robust feature of the simulation and are not simply a transient
phenomenon.   

\subsection{The effects of varying $\zeta_0$}

\begin{figure}
\begin{center}
\epsfxsize\hsize\epsffile{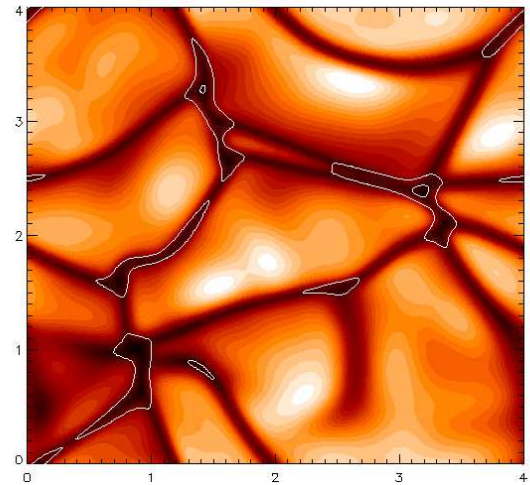}
\caption{The magnetic flux distribution during the kinematic phase for
  $\zeta_0=1.2$. Like Fig.~2, this plot is for $t=0.24$ and shows
  the spatial variation of the temperature and the vertical magnetic
  field component in a horizontal plane at $(z/d)=0.05$ (i.e. near the
  upper surface). For ease of comparison, the same scales have been
  used in Fig.~2 and Fig.~5 for both the magnetic field and the
  temperature.\label{fig:5}}  
\end{center}
\end{figure}

One of the key parameters in these simulations is the magnetic
Reynolds number, $Rm$. Computational restrictions limit the range of
values of $Rm$ that can be considered, and all values that can be
simulated will be very much smaller than real photospheric
values. However, the simulation that has already been described
qualitatively illustrates some of the main physical processes that
occur during photospheric magnetic flux amplification. In this
Section, we assess the effects of repeating this simulation for
different values of $\zeta_0$. This is equivalent to varying the
magnetic Reynolds number of the flow, which is inversely proportional
to $\zeta_0$: smaller values of $Rm$ correspond to larger values of
$\zeta_0$ and vice-versa. In order that comparisons can easily be made
between the different cases, all magnetohydrodynamical simulations are
started from exactly the same hydrodynamic initial conditions.  

\begin{table}
\begin{center}
\caption{The $\zeta_0$-dependence of the magnitude of the vertical
  magnetic field component at a fixed position at the upper surface of the
  computational domain. The time is fixed at $t=0.12$, and the centre
  of the magnetic feature is at
  $\left(x,y\right)=\left(3.16,2.39\right)$ in all cases. Magnetic
  field strengths are normalised in terms of the strength of the
  imposed magnetic field. Also shown (in the middle column) is the
  corresponding value of $Rm$ for each value of $\zeta_0$.\label{table:1}} 
\begin{tabular}{@{}lcc}
\hline $\zeta_0$ & $Rm$ & $B_z (max)/B_z (initial)$ \\ \hline 0.2 &
117 & 3.61 \\
0.3 & 83 & 3.50 \\ 0.4 & 62 & 3.40 \\ 0.6 & 42 & 3.23 \\ 1.2 & 22 &
2.85 \\ 2.4 & 11 & 2.42 \\ \hline
\end{tabular}
\end{center}
\end{table}

\begin{figure*}
\begin{center}
\epsfxsize\hsize\epsffile{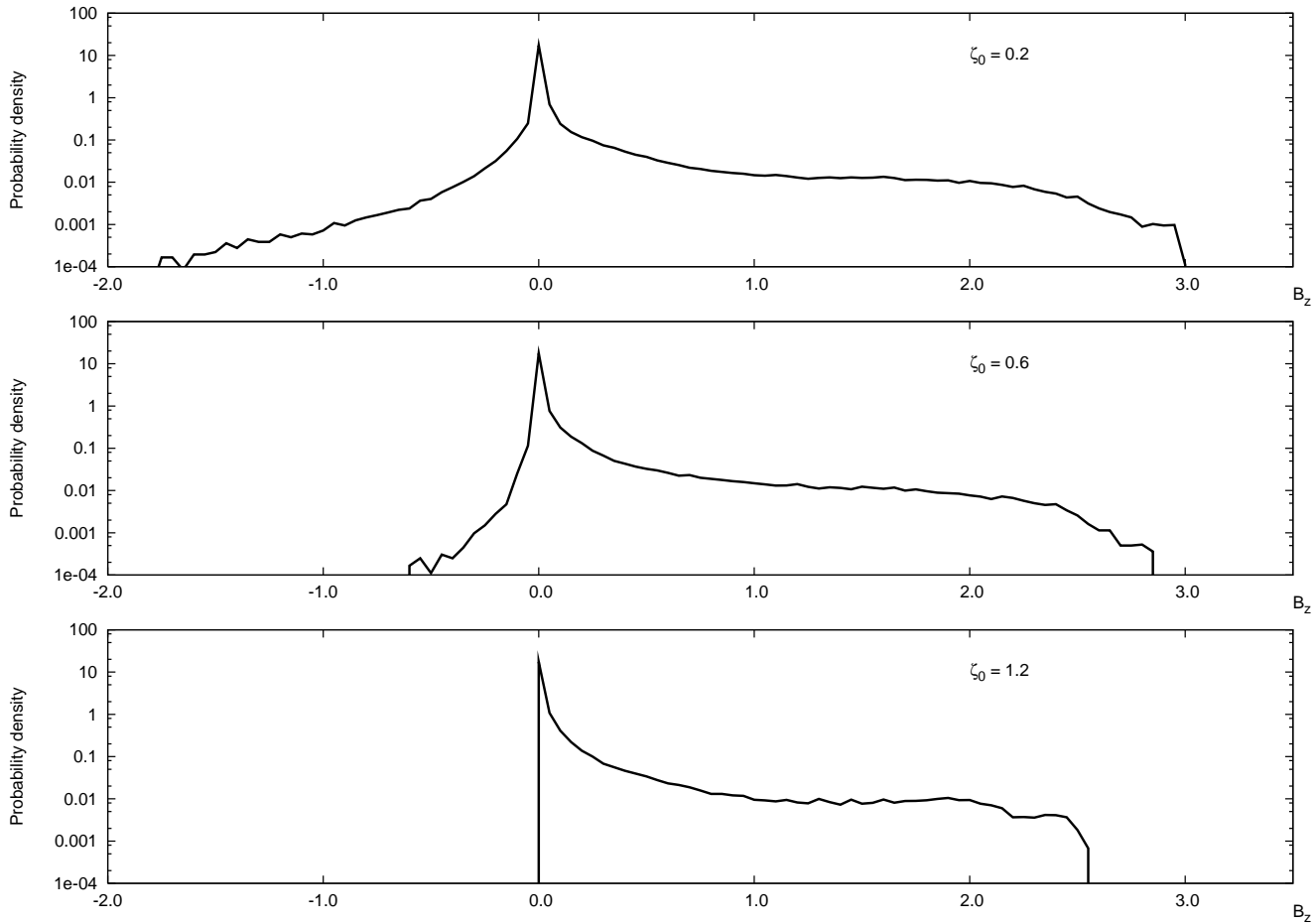}
\caption{Probability density functions for the vertical component of
  the magnetic field at the upper surface of the computational
  domain. These functions correspond to different values of
  $\zeta_0$. These values are $\zeta_0=0.2$ (top), $\zeta_0=0.6$ (middle) and
  $\zeta_0=1.2$ (bottom). For smaller values of $\zeta_0$ (or,
  equivalently, larger values of $Rm$) there is a higher probability
  of stronger fields and there is a greater proportion of reversed
  polarity magnetic flux.\label{fig:6}} 
\end{center}
\end{figure*}

\par Even during the brief kinematic phase, clear trends are observed
as the parameter $\zeta_0$ is varied. The effects of some of these
trends are illustrated in Fig.~\ref{fig:5}, for $\zeta_0=1.2$. Like
Fig.~\ref{fig:2}, this shows the spatial distribution of the
temperature and the vertical component of the magnetic field near the
upper surface of the computational domain. In order to allow direct
comparison between Fig.~\ref{fig:2} and Fig.~\ref{fig:5}, this
snapshot of the simulation is also taken at $t=0.24$, and the contours
are plotted at the same levels -- this implies that the scales for
temperature and magnetic field correspond directly to those used in
Fig.~\ref{fig:2}. In this kinematic phase, the temperature field is
the same in both plots; however it is immediately apparent that the
magnetic field structures in Fig.~\ref{fig:5} are weaker and less
localised than those seen in the $\zeta_0=0.2$ case. This is a
consequence of the increased importance of diffusion for larger values
of $\zeta_0$ (equivalently, lower values of $Rm$). The influence that
$\zeta_0$ has upon the magnetic field in this kinematic phase is quantified in
Table~\ref{table:1}. Here, we measure the strength of the vertical
component of the magnetic field at
$\left(x,y,z\right)=\left(3.16,2.39,0\right)$ and $t=0.12$, as a
function of $\zeta_0$. This spatial location corresponds to the centre
of the strongest surface magnetic feature at $t=0.12$, which does not
depend upon $\zeta_0$ in the kinematic regime (although the location
of this peak field is obviously a function of time). The table confirms
that, at a fixed time, there are weaker fields at lower values of
$Rm$. For simple flows, it is possible to construct a similarity
solution for this kind of kinematic flux concentration, which leads to
a power-law relationship between the peak magnetic field and the
magnetic Reynolds number (see, e.g., Proctor \& Weiss,
1982). Unfortunately, due to the complexity of the flow patterns in
this case, there appears to be no analogous scaling here.  

\par For $\zeta_0=0.2$, it was found that the high magnetic pressure in the
strongest magnetic field concentrations quickly led to the partial
evacuation of these magnetic features. Since these magnetic flux
concentrations grow more gradually for larger values of $\zeta_0$,
this partial evacuation process is much slower in these
cases and the effects of diffusion tend to play a more dominant role
in limiting the flux concentration. Therefore, the amount of evacuation
that occurs decreases with increasing $\zeta_0$. This demonstrates that the
rapid intensification that occurred in the $\zeta_0=0.2$ case is very
much a high magnetic Reynolds number phenomenon. However, some
evacuation is observed in all cases that were investigated except
$\zeta_0=2.4$, and wherever features do become partially evacuated,
super-equipartition magnetic features are observed. Fig.~\ref{fig:6}
shows time-averaged probability density functions (pdfs) for the
vertical component of the surface magnetic field, for three different
values of $\zeta_0$. In all cases, the pdfs peak at $B_z=0$, which
implies that the majority of the domain is field-free. In addition,
there is a significant component of reversed magnetic flux at lower
values of $\zeta_0$. At higher $Rm$ there is a greater tendency for
flux to be advected with any fluid flow, and these convective motions
certainly have the potential to reverse magnetic flux at the granular
boundaries at the surface of the domain. Most interestingly, there is
not a large difference between the peak fields in the pdfs for
$\zeta_0=0.6$ and $\zeta_0=0.2$. Equivalently, the magnitude of the
peak field appears to be only weakly-dependent upon $Rm$ in this
particular parameter regime. This suggests that rather than being
controlled by diffusive effects, the peak attainable magnetic field in
this parameter regime is determined by the combined effects of the
external gas and dynamic pressures. So, although ultra-intense magnetic
features form more rapidly at higher $Rm$, these features can also be
produced at lower values of $Rm$, provided that magnetic diffusion
does not completely inhibit the flux concentration process before the
magnetic field can become dynamically active.   

\section{Conclusions}

\par This paper describes results from a series of numerical
experiments that were designed to investigate the formation of
localised magnetic flux concentrations at the solar photosphere. By
adopting an idealised model, it was possible to assess the effects
that varying the magnetic Reynolds number might have upon this flux
intensification process. As expected, magnetic flux tends to
accumulate preferentially in convective downflows, where it forms
localised features -- the horizontal scale of these features decreases
with increasing magnetic Reynolds number. High magnetic pressures lead
to the partial evacuation of these features as they form. At high
values of $Rm$, the resulting field strengths are typically much
larger than the equipartition value at which the local magnetic energy
density balances the mean kinetic energy density of the surrounding
granular convection. In addition, the strongest magnetic fields that
are formed in the upper layers of the domain exert a magnetic pressure
that is significantly larger than the external gas pressure. The
appearance of these ultra-intense magnetic fields shows that the
dynamic pressure that is associated with the surrounding convection
must be playing a key role in the confinement of these magnetic
features to localised regions. Some super-equipartition magnetic
fields are found in all cases except the most diffusive case
(corresponding to $\zeta_0=2.4$).  

\par It is interesting to relate these results to convective
collapse models (see, e.g., Webb \& Roberts 1978; Spruit \& Zweibel
1979). Although those models are certainly an
idealised representation of photospheric magnetic field
intensification, our simulations suggest that they correctly
identify the most important process in the formation of
super-equipartition magnetic fields, namely the partial evacuation of
magnetic regions by convective downflows. Our simulations do, however, 
raise some important issues relating to convective
collapse. First, convective collapse models typically adopt an initial
condition that corresponds to a thin flux tube embedded in a static
atmosphere. Since such magnetic features form in well-established
convective downflows, this static equilibrium never occurs in our
simulations. Additionally, the evacuation process seems to begin before the
flux accumulation process finishes. So our simulations could be
seen as describing an ``adjustment'' rather than an instability. The
second point that is raised by these simulations concerns the pressure
balance. Convective collapse models assume a constant balance between
the external gas pressure and the gas and magnetic pressures within
the flux tube. Our simulations indicate that the dynamical effects of
the surrounding convection are playing a significant role in the
confinement of these magnetic features. The neglect of this dynamic
pressure underestimates the strength of the strongest magnetic
fields that can be generated. Finally, Cameron \&
Galloway (2005) have modelled aspects of convective collapse by  conducting numerical simulations of laminar magnetoconvection in a simplified geometry. From these simulations they argue that super-equipartition magnetic features must be structured on a
kinematic scale at the solar photosphere. Although only a limited
range of values for $Rm$ has been considered here, our simulations
do produce some super-equipartition fields on larger scales, and so
contradict their view. 

\par When comparing results from these simulations with photospheric
observations, it is important to remember some of the simplifying
assumptions in this model. Processes such as radiative
transfer have been neglected, so it is not possible to make detailed spectral
comparisons between these simulations and observations. Nevertheless, this
idealised approach has had a great deal of success in reproducing
qualitative features of photospheric magnetoconvection. In qualitative
terms, our simulations appear to be consistent with solar
observations and provide a plausible explanation for the appearance
of super-equipartition magnetic features in the intergranular lanes in
the quiet Sun. However, this model does have other limitations, notably
the fact that many of the parameters that are used are not closely related
to realistic solar values. We have also used highly idealised boundary
conditions in these simulations: for example we impose a rigid
boundary at the upper surface. It should be stressed that, although
the transition to a subadiabatic stratification in the photosphere
provides a ``softer'' boundary condition, this will not prevent the
formation of extremely strong magnetic fields. Another
limitation that should be noted is the fact that (due to the periodic
boundary conditions) the imposed magnetic flux is independent of both
$z$ and $t$. This implies that net vertical magnetic flux cannot enter
or leave the domain, therefore the initial non-zero magnetic flux is
an invariant quantity in the model. In the quiet Sun, there is a
continuous emergence of mixed polarity magnetic flux, which will
interact with existing magnetic features. Such interactions will tend
to limit the lifetimes of these magnetic features. This is something
that cannot be represented in our idealised simulations, though it may
be possible to make progress by careful choice of initial conditions.  
 
\par In the Introduction, we noted that it is still not clear
whether quiet Sun magnetic features are simply fragments of reprocessed
magnetic flux or whether they are generated locally as the result of
small-scale dynamo action. In fact, this model of compressible
convection can drive a small-scale dynamo once the magnetic Reynolds number
exceeds a threshold of about $Rm=250$. However, even for a
marginally-excited dynamo, the magnetic Prandtl number is of order $2$
which is very much larger than the magnetic Prandtl number in the solar
photosphere. Whether or not such a dynamo could operate in the low
magnetic Prandtl number regime has been the subject of
considerable debate (see, for example, Boldyrev \& Cattaneo 2004;
Schekochihin et al. 2005). Although our idealised simulations do assume
some pre-existing magnetic field (i.e. they do not generate this field
self-consistently), the processes of magnetic flux expulsion and
intensification that are illustrated by these simulations are generic
and are therefore likely to be of relevance to the solar photosphere
whether or not a local dynamo is operating.   

\par This work is motivated by high resolution observations of
the solar photosphere. Although current ground-based instruments, such
as the Swedish 1-metre Solar Telescope, are already provided very
detailed images of the photosphere, it is likely that newer
instruments (including those carried on the recently launched Hinode
satellite) will reveal many new features of photospheric
magnetoconvection. Over the next few years, these new observations
will enable us to refine our current theoretical models, but will also
inevitably present new theoretical challenges.

\section*{Acknowledgements}

This work was supported by PPARC/STFC while PJB held a postdoctoral
appointment at DAMTP in Cambridge. The numerical simulations that
were described in this paper made use of computing facilities belonging to the
UKMHD Consortium (based at the University of St Andrews) and the
Cambridge-Cranfield High Performance Computing Facility. We would also
like to thank Fran\c{c}ois Rincon and the referee for their helpful comments.

\bsp

\label{lastpage}

\end{document}